\documentclass[a4paper,12pt]{article}

\usepackage{ifpdf}

\newif\ifpdf
\ifx\pdfoutput\undefined
  \pdffalse
\else
  \pdfoutput=1
  \pdftrue
\fi

\RequirePackage{xspace} %
\RequirePackage{subfigure} %
\RequirePackage[centertags]{amsmath} %
\RequirePackage{amssymb}
\RequirePackage{wrapfig} %
\RequirePackage{calc} %
\RequirePackage{ifthen}
\RequirePackage{tabularx} %
\RequirePackage{flafter} %
\RequirePackage{fancyhdr} %

\ifpdf
  \RequirePackage[pdftex]{color}%
  \RequirePackage{colortbl}%
  \RequirePackage{array}%
  \RequirePackage[pdftex]{graphicx}

  \RequirePackage[ pdftex, plainpages = false, pdfpagelabels,
                 pdfpagelayout = useoutlines,
                 bookmarks,
                 breaklinks = true,
                 linktocpage,
                 pagebackref,                      
                 colorlinks = true,
                 linkcolor = blue,
                 urlcolor  = blue,
                 citecolor = blue,
                 anchorcolor = blue,
                 hyperindex = true,
                 hyperfigures
                 ]{hyperref}

\else
  \RequirePackage{color}
  \RequirePackage{colortbl}
   \RequirePackage{array}
  \RequirePackage[dvips]{graphicx}
  \RequirePackage{hyperref}
  \usepackage{rotating}
\fi

\usepackage{makeidx} 
\usepackage{setspace} 
\usepackage{rotating} 
\usepackage{ecltree}
\usepackage{epic}
\usepackage{supertabular}  
\usepackage{color}
\usepackage{exscale}
\usepackage{fontenc}
\usepackage{ifthen}
\usepackage{latexsym}
\usepackage{makeidx}
\usepackage{syntonly}
\usepackage{inputenc}
\usepackage{graphicx}
\usepackage{setspace}
\usepackage{caption2}
\usepackage[english]{babel}
\usepackage[square, comma,numbers,sort&compress]{natbib}
\usepackage{hypernat}
\usepackage{boxedminipage}
\usepackage{framed}
\usepackage{longtable}
\usepackage[all]{hypcap}    
\usepackage{algorithm2e}
\usepackage{algorithmic}
\usepackage{lscape}
\usepackage{pdflscape}

\newcommand{\note}[1]{\marginpar[left]{\singlespace \tiny #1}}

\renewcommand{\sectionmark}[1]%
      {\markright{\thesection\ #1}} 

\renewcommand{\note}[1]{}

\onehalfspace 

\begin{document}
\begin{center}
{\Large Solving the flow fields in conduits and networks using energy minimization principle with
simulated annealing}
\par\end{center}{\Large \par}

\begin{center}
Taha Sochi
\par\end{center}

\begin{center}
{\scriptsize University College London, Department of Physics \& Astronomy, Gower Street, London,
WC1E 6BT \\ Email: t.sochi@ucl.ac.uk.}
\par\end{center}

\begin{abstract}
\noindent In this paper, we propose and test an intuitive assumption that the pressure field in
single conduits and networks of interconnected conduits adjusts itself to minimize the total energy
consumption required for transporting a specific quantity of fluid. We test this assumption by
using linear flow models of Newtonian fluids transported through rigid tubes and networks in
conjunction with a simulated annealing (SA) protocol to minimize the total energy cost. All the
results confirm our hypothesis as the SA algorithm produces very close results to those obtained
from the traditional deterministic methods of identifying the flow fields by solving a set of
simultaneous equations based on the conservation principles. The same results apply to electric
ohmic conductors and networks of interconnected ohmic conductors. Computational experiments
conducted in this regard confirm this extension. Further studies are required to test the energy
minimization hypothesis for the non-linear flow systems.

\vspace{0.3cm}

\noindent Keywords: fluid dynamics; pressure field; tube; network; porous media; energy
minimization; simulated annealing; stochastic method; electric conductor; electric network.

\par\end{abstract}

\begin{center}

\par\end{center}

\section{Introduction} \label{Introduction}

One of the fundamental physical principles that regulate Nature's behavior is optimization which
reflects a prejudice that leads to minimizing or maximizing selected physical quantities. For
instance, Nature has a tendency to maximize the entropy of dynamic systems, as given by the second
law of thermodynamics, but to minimize the passage time of light as summarized by Fermat's least
time principle. Many physical laws have been deduced or derived from optimization arguments and
hence it is one of the main pillars of modern science. As a result, extensive branches of
mathematical and computational disciplines have been developed to deal with modeling and
quantifying optimization problems with widespread applications in physical and social sciences.

One of the powerful and widely used optimization methods is simulated annealing
\cite{MetropolisRRTT1953, KirkpatrickGV1983, Cerny1985} which is a stochastic computational
technique based on the physical principles of statistical mechanics. The essence of this method is
to emulate the process of controlled and slow cooling of liquified substances so that they reach
their minimal energy configuration in their solid state. The main advantages of simulated annealing
are its simplicity and wide applicability to large classes of optimization problems as well as its
high success rate of avoiding traps of local minima which other deterministic and stochastic
methods are more likely to fall in. Furthermore, in many cases it is the only viable method as
combinatorial enumeration and other analytical or conceptually-based methods are not viable or
available in those circumstances.

The main disadvantage of simulated annealing is its generally high computational cost in terms of
CPU time. Although this is true for commonplace problems where alternative methods are available,
in some cases simulated annealing is more efficient even in terms of CPU time when the
computational cost grows exponentially and hence the cost of traditional methods, assuming their
viability, becomes much higher than the cost of SA. In fact this is one of the main reasons why
simulated annealing and similar stochastic methods are invented and widely used in all sorts of
optimization problems as can be inferred, for instance, from the number of citations of the SA
founding papers \cite{MetropolisRRTT1953, KirkpatrickGV1983, Cerny1985}. Regardless of this,
nowadays the computational cost is a trivial factor in many cases considering the huge advances
over the last few decades in the hardware and software development and the availability of
multi-processor platforms, even for personal use, with relative ease of parallelization.

In the present paper we suggest and examine a hypothesis that the driving field, like pressure and
potential difference, in conducting elements and networks of interconnected elements will adjust
itself to minimize the energy cost of transporting a given quantity of fluid through the transport
device. We use simulated annealing with supporting arguments to achieve the minimization objective
and establish the energy minimization hypothesis. We restrict our attention in the current
investigation to the linear transport systems where the driving and induced fields are linearly
correlated, such as the flow of Newtonian fluids in rigid tubes and networks of interconnected
rigid tubes and the flow of electric current in ohmic components and networks of interconnected
ohmic components, although we will briefly discuss some issues related to the non-linear systems
for the sake of completeness. The case of non-linear flow systems requires further investigation to
reach definite conclusions although initial examination indicates that they are also subject to the
energy minimization rules.

The model that is used in the current investigation to present the energy minimization principle
and elaborate the SA optimizing algorithm is from fluid mechanics, namely the aforementioned
example of the flow of Newtonian fluids in rigid tubes and networks. However, since this flow model
is mathematically equivalent to the flow model of electric current in ohmic devices, our
investigation and conclusions will naturally extend to this case as well.

\section{Method}

In this study we assume a laminar, incompressible, isothermal, time-independent, pressure-driven,
fully-developed flow with minor entry and exit effects. Our plan for establishing the energy
minimization principle through solving the flow fields by simulated annealing is summarized in the
following points

\begin{enumerate}

\item
We establish the existence and uniqueness of the flow solution in general.

\item
We explain how to adapt simulated annealing to find a flow solution in single conduits and networks
of interconnected conduits.

\item
We demonstrate that the solution found by simulated annealing is a correct one, it minimizes the
total energy of fluid transportation, and this minimum is a global rather than a local one.

\end{enumerate}


As for the {\bf first point}, the existence and uniqueness of flow solutions for single conduits is
a thorny issue from the theoretical and mathematical viewpoint. However, it can be established by
the forthcoming physical argument which we presented in the context of network discussion.
Regardless of this, we can take this for granted by claiming it is an intuitive assumption. In fact
most of the ongoing studies in this field should be based explicitly or implicitly on such an
assumption especially the purely theoretical ones which are not supported by experimental or
observational evidence.

With regard to networks, the existence and uniqueness conditions can be established by the
following argument assuming the existence and uniqueness of the solution on their individual
conduits. For the linear case, to describe such flow networks we set a system of $N$ simultaneous
linear equations in $N$ unknowns where $N$ is the total number of nodes in the network which
includes the boundary as well as the internal nodes. The equations of the inlet and outlet nodes
are derived from the boundary conditions while the equations of the internal nodes are derived from
the mass conservation principle in conjunction with the characteristic flow relation that
correlates the driving and induced fields, like $p$ and $Q$ in Hagen-Poiseuille law. Since these
equations are linearly independent, due to the fact that no two equations share the same sequence
of conducting elements and hence they cannot be represented as scalar multiples of each other, we
have a system of $N$ linearly independent equations in $N$ unknowns and hence a solution does {\it
exist} and it is {\it unique} according to the rules of algebra.

With regard to the non-linear systems, there is no general condition that guarantees the existence
or uniqueness of solution. However, physical rather than mathematical arguments can be proposed to
establish the existence and uniqueness of solutions even for the non-linear systems in case such an
extension is required. It can be argued that for both the linear and non-linear systems a solution
should exist and it should be unique regardless of all these elaborate mathematical considerations
because as long as our mathematical models reflect the essential features of the reality of these
classical deterministic systems, the soundness and accuracy of these models will guarantee the
existence and uniqueness conditions. Such line of reasoning should be sufficient for the purpose of
establishing our energy minimization argument since we have no interest in those theoretical and
mathematical subtleties.


As for the {\bf second point}, the time rate of energy consumption, $I$, of fluid transport through
a conducting device considering the type of flow systems that meet our stated assumptions, is given
by

\begin{equation}
I=\Delta p\,Q
\end{equation}
where $\Delta p$ is the pressure drop across the conducting device and $Q$ is the volumetric flow
rate of the transported fluid. For a single conduit that is discretized into $n$ sections indexed
by $i$, the total energy consumption rate, $I_t$, is given by

\begin{equation} \label{It}
I_t=\sum_{i=1}^n\Delta p_i Q_i
\end{equation}
A similar equation applies for a network of $n$ interconnected conduits, with $n$ standing for the
number of conduits rather than sections, if only the nodal pressure values are required. If the
axial pressure values at the midpoints of conduits are also required, a discretization scheme,
similar to the one used for single conduits, can be used where it is needed. However, for the
linear systems, only nodal pressures are necessary to compute since the midpoint values can be
obtained by a simple linear interpolation scheme.

The role of the simulated annealing protocol then is to find the set of pressure values, $p_k$
where $k$ indexes the discretization and nodal points, that minimizes the rate of total energy
consumption, given by Equation \ref{It}. Accordingly, the $p_k$ values can be freely adjusted by
the SA routine to satisfy the minimization requirement. The exception to this is the boundary
values which are held constant in all SA iterations to satisfy the imposed boundary conditions. As
for the $Q_i$ values, they are computed in each SA iteration from the analytical expression of the
model that correlates the volumetric flow rate to the pressure drop, i.e. Hagen-Poiseuille law in
the case of Newtonian flow systems, using the most recent values of $p_k$. Unlike the traditional
solution methods, such as the residual-based Newton-Raphson technique, no conservation principle,
such as mass continuity, is needed in the proposed SA solution scheme.

Regarding our simulated annealing algorithm, we use a standard scheme as described by many papers
and monographs written on simulated annealing and stochastic methods. Briefly, we start by
initializing the pressure values, $p_k$, randomly except the boundary ones which are set to the
values required by the imposed boundary conditions. We then compute the cost function, which is
initially set to a very high value, as given by Equation \ref{It}. The new pressure field solution,
as defined by the set of $p_k$ values, is then accepted if the cost function obtained from the
current iteration is less than the cost function of the most recently accepted solution. If the
cost function of the new solution is higher, it may also be accepted but with a probability $P$
given by

\begin{equation}
P=e^{\frac{I_{t_{m-1}}-I_{t_m}}{T_c}}
\end{equation}
where $m$ is an index for the current iteration and $T_c$ is the current value of the annealing
parameter (temperature) which we assume to have the same units as $I$. The $p_k$ values are then
adjusted randomly using a random number generator and the cycle is repeated. The annealing control
parameter, $T$, is decremented persistently as the annealing goes on. The algorithmic procedure
will terminate when the annealing control parameter reaches its lower limit which is normally set
to virtually zero, and hence the final solution will be taken as the last accepted set of $p_k$
values according to the minimization criterion.

By finding the pressure field, the volumetric flow rate field will be easily obtained from the
analytical expression of the flow in single conduits that links $Q$ to $\Delta p$. For networks,
the total inflow/outflow can be obtained by computing and summing the volumetric flow rates of the
inlet/outlet boundary conduits.

As well as implementing this standard SA scheme, we experimented with a number of similar SA
schemes which differ in elaboration and complexity, for instance by the way that $T$ varies or by
the accepted exit condition from the interim cycles, but since they all produce very similar
results we decline to include these irrelevant details.


As for the {\bf third point}, the verification of the single conduit solution is trivial since the
axial pressure varies linearly with the conduit axial coordinate. For the networks, the obtained
solution from the simulated annealing procedure can be checked for correctness by simultaneously
satisfying the analytical flow relation that links $Q$ to $\Delta p$ on each conduit plus the mass
conservation principle on each internal node. Detailed explanations about these verification
conditions and other related issues are given in \cite{SochiPois1DComp2013}. The simulated
annealing solution can also be verified more easily by comparison to the solution found by other
methods, mainly the deterministic ones such as the residual-based Newton-Raphson scheme which
originates from the existence and uniqueness proof that we outlined earlier. Further details about
this can also be found in \cite{SochiPois1DComp2013}.

Since simulated annealing is essentially a minimization algorithm, the minimal energy principle is
established by finding the optimal solution so no further proof is required to establish the fact
that the obtained solution is not a maximum or a stationary inflection point, unlike solutions
obtained by other methods where such a proof may be needed. Since this minimizing solution is the
only possible solution, as established by the uniqueness condition, it should be a global rather
than a local minimum since no other solution, minimizing or non-minimizing, is possible.

\section{Results}

\begin{figure}[!h]
\centering{}
\includegraphics
[scale=0.6] {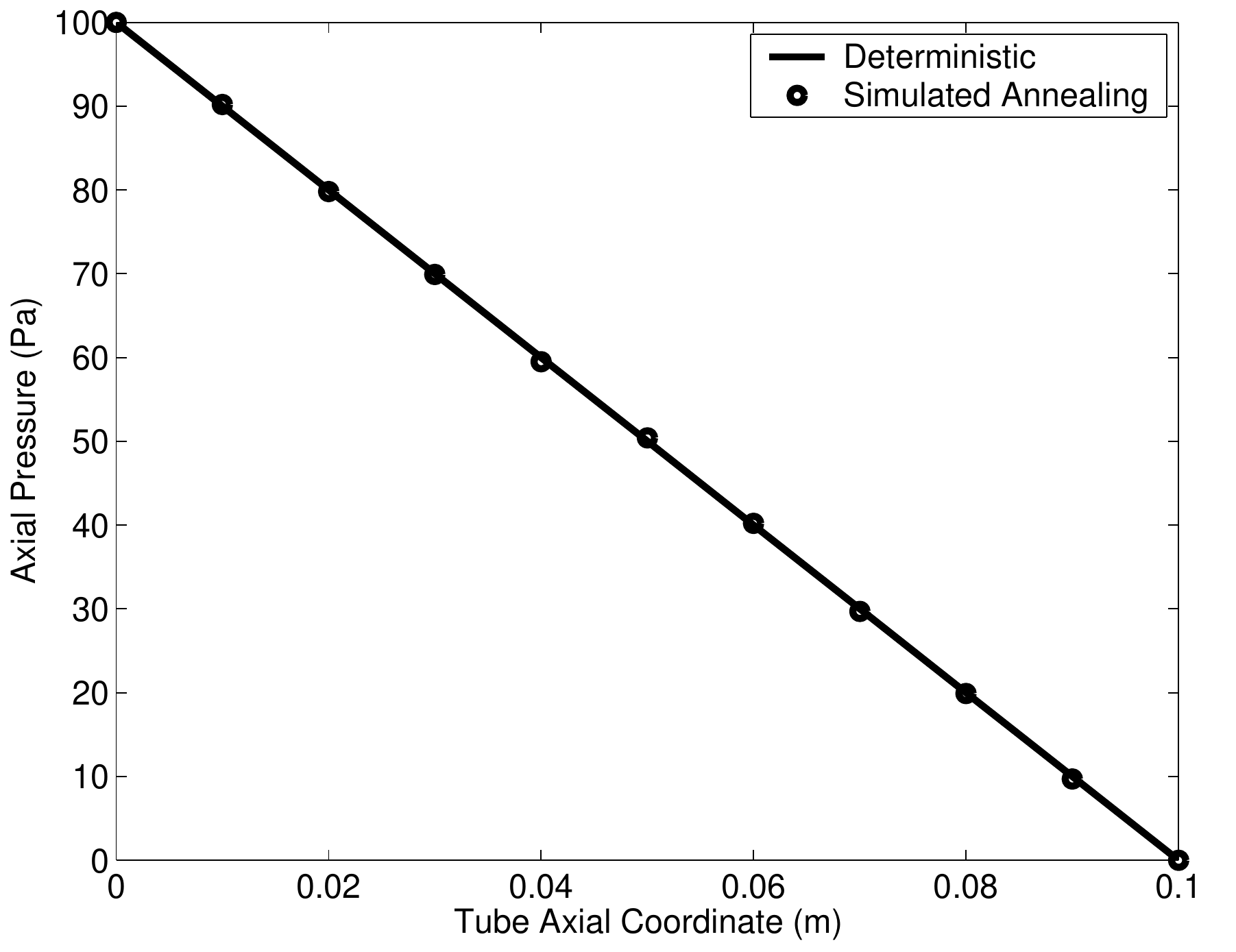} \caption{Comparison between the pressure field solution, as a function of
the axial coordinate for a single tube, as obtained from the simulated annealing method based on
the energy minimization principle and the solution obtained by interpolation or from the classical
deterministic residual-based Newton-Raphson method.} \label{TubePX}
\end{figure}

\begin{figure}[!h]
\centering{}
\includegraphics
[scale=0.6] {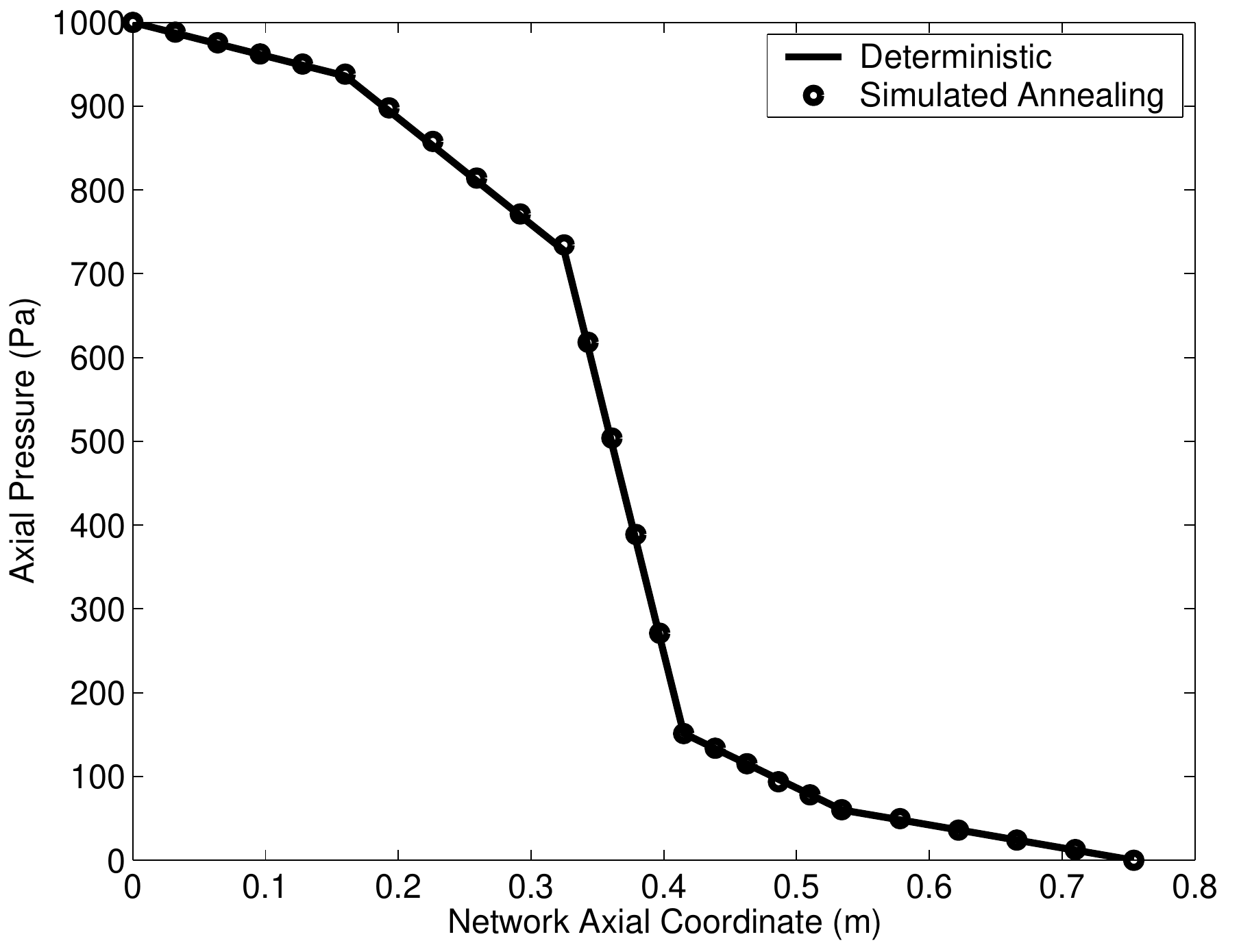} \caption{Comparison between the pressure field solution, as a function
of the axial coordinate for a one-dimensional network of five serially-connected tubes with
different lengths and radii, as obtained from the simulated annealing method based on the energy
minimization principle and the solution obtained from the classical deterministic residual-based
Newton-Raphson method.} \label{LinearNetPX}
\end{figure}

The simulated annealing algorithm, which is described in the last section, was implemented in a
computer code and results were obtained and analyzed using various models of fluid transport
devices including rigid tubes, one-, two- and three-dimensional networks of interconnected rigid
tubes. The multi-dimensional networks include fractal, cubic and orthorhombic lattice models; the
detailed description of these networks is given, for instance, in \cite{SochiPois1DComp2013,
SochiPoreScaleElastic2013}. These model tubes and networks vary in their size, geometry,
connectivity, number of nodes and segments, and statistical distributions. Various boundary
conditions and discretization schemes with different model Newtonian fluids were also employed in
these computations. A sample of these results is given in Figures \ref{TubePX} and
\ref{LinearNetPX} and Table \ref{TempTableCPD}. All the results represent averaged data over many
similar runs where no run has failed to meet the rigorous criteria set for accepting the solutions.

As seen, the SA results agree very well with the verified deterministic solutions. One thing that
should be highlighted is the general trend of increasing the size of the errors with increasing the
size of the networks. The main reason is our tendency to reduce the required CPU time and hence the
results can be improved if longer CPU time is allowed and more elaborate simulated annealing
schemes are employed. The size of the errors in these results, nonetheless, does not affect the
definite conclusion that can be drawn from this investigation about the validity of the energy
minimization principle as a governing rule in determining the flow fields ($p$ and $Q$) in the
fluid transport systems. After all, such errors, and even larger ones, are expected to contaminate
the results obtained by stochastic methods. In fact significant errors can occur even in the
numerical deterministic methods although the size of the errors in the latter is usually less than
that in the former.

These results do not only establish the energy minimization principle but also establish a novel
way for solving the flow fields in conduits and networks by stochastic methods in general and by
simulated annealing in particular. Most of the reported results are obtained within seconds or
minutes of CPU time on a normal laptop computer where the length of the CPU time mainly depends on
the number of discretized elements in the transport device assuming the employment of the same SA
parameters and scheme. Although the proposed stochastic method is generally slower than the
traditional deterministic methods, it may be possible to compete with the traditional deterministic
methods even in speed for very large networks especially if more elaborate SA schemes are employed.
A big advantage, however, of the proposed stochastic method is the trivial memory cost; a factor
based on its serial nature and the redundancy of employing a numerical solver that is usually
needed in the deterministic methods. Anyway, the proposed stochastic method is a tool that is
advantageous to be available to scientists and engineers and it definitely can have useful
applications in some exceptional circumstances at least.

\clearpage

\begin{table} [!h]
\caption{Statistical distribution parameters of the percentage relative difference of the nodal
pressures between the Newton-Rapson deterministic solutions and the simulated annealing solutions
for a number of 2D fractal (F) and 3D orthorhombic (O) networks with the given number of segments
(NS) and number of nodes (NN). The results represent averaged data over multiple runs for each
network. The meaning of the statistical abbreviations are: Min for minimum, Max for maximum, SD for
standard deviation, and Avr for average. \label{TempTableCPD}}
\begin{center} 
\begin{tabular*}{\textwidth}{@{\extracolsep{\fill}}cccccccc}
\hline
{\bf Index} & {\bf Type} &   {\bf NS} &   {\bf NN} &  {\bf Min} &  {\bf Max} &   {\bf SD} &  {\bf Avr} \\
\hline
         1 &          F &         15 &         16 &      -0.42 &       0.35 &       0.15 &      -0.06 \\

         2 &          F &         31 &         32 &      -0.70 &       0.67 &       0.22 &      -0.16 \\

         3 &          F &         63 &         64 &      -0.79 &       0.75 &       0.24 &      -0.05 \\

         4 &          F &        127 &        128 &      -0.77 &       0.85 &       0.27 &       0.19 \\

         5 &          F &        255 &        256 &      -1.29 &       1.15 &       0.25 &       0.07 \\

         6 &          F &        511 &        512 &      -1.87 &       1.74 &       0.35 &      -0.06 \\

           &            &            &            &            &            &            &            \\

         7 &          O &         72 &         45 &      -0.51 &       0.58 &       0.20 &       0.10 \\

         8 &          O &         99 &         60 &      -0.89 &       0.78 &       0.22 &      -0.11 \\

         9 &          O &        136 &         80 &      -1.11 &       1.12 &       0.26 &       0.23 \\

        10 &          O &        176 &         96 &      -1.42 &       1.37 &       0.32 &      -0.17 \\

        11 &          O &        216 &        112 &      -1.65 &       1.57 &       0.37 &      -0.29 \\

        12 &          O &        275 &        140 &      -1.74 &       1.68 &       0.44 &       0.35 \\
\hline
\end{tabular*}
\end{center}
\end{table}

\section{Conclusions} \label{Conclusions}

We demonstrated, through the use of simulated annealing, the validity of the assumption of energy
minimization principle as a governing rule for the flow systems in the context of obtaining the
driving and induced fields in single conduits and networks of interconnected conduits where the
driving and induced fields are linearly correlated, e.g. the flow of Newtonian fluids through rigid
tubes and networks or the flow of electric current through ohmic devices. All the results support
the proposed energy minimization principle.

There are two main outcomes of this investigation. First, a novel method for solving the flow
fields that is based on a stochastic approach is proposed as an alternative to the traditional
deterministic approaches such as the residual-based Newton-Raphson and finite element methods.
Although this new numerical method may not be attractive in most cases where it is outperformed in
speed by the traditional methods, it can surpass the other methods in other cases and may even be
the only viable option in some circumstances where the flow networks are very large.

The second outcome, which in our view is the most important one, is the theoretical conclusion that
energy minimization principle is at the heart of the flow phenomena and hence it governs the
behavior of flow systems; the linear ones at least. This is inline with our previous investigations
\cite{SochiVariational2013, SochiSlitPaper2014} which are based on minimizing the total stress in
the flow conduits. The current investigation may be extended in the future to the non-linear case.

\vspace{1cm}
\phantomsection \addcontentsline{toc}{section}{Nomenclature} %
{\noindent \LARGE \bf Nomenclature} \vspace{0.5cm}

\begin{supertabular}{ll}
$I$                     &   time rate of energy consumption for fluid transport \\
$I_t$                   &   time rate of total energy consumption for fluid transport \\
$P$                     &   probability of accepting non-minimizing SA solution \\
$p$                     &   pressure \\
$\Delta p$              &   pressure drop across flow conduit \\
$Q$                     &   volumetric flow rate \\
$T$                     &   annealing control parameter (temperature) \\
$T_c$                   &   current value of annealing control parameter \\
\end{supertabular}

\vspace{1cm}
\phantomsection \addcontentsline{toc}{section}{References} %
\bibliographystyle{unsrt}

\end{document}